\documentclass[3p,times]{elsarticle}

\usepackage{ecrc}
\usepackage{wrapfig}
\RequirePackage{lineno}

\volume{00}

\firstpage{1}

\journalname{Nuclear Physics A}

\runauth{}

\jid{nupha}

\usepackage{amssymb}
\usepackage[figuresright]{rotating}

\newcommand{\sqn}{\mbox{$\sqrt{s_{NN}}$}}

\newcommand{\Npart}{\mbox{$N_{\mathrm{part}}$}}
\newcommand{\Ncoll}{\mbox{$N_{\mathrm{coll}}$}}

\newcommand{\pp}{\mbox{p+p}}
\newcommand{\sumet}{\mbox{$\sum E_T$}}
\newcommand{\rpc}{\mbox{$R_{\mathrm{PC}}$}}

\newcommand{\pT}{\mbox{$p_{\mathrm{T}}$}}
\newcommand{\pt}{\mbox{$p_{\mathrm{T}}$}}
\newcommand{\et}{\mbox{$E_{\mathrm{T}}$}}

\newcommand{\zee}{\mbox{$Z\rightarrow e^{+}e^{-}$}}
\newcommand{\zmm}{\mbox{$Z\rightarrow\mu^{+}\mu^{-}$}}

\newcommand{\ptZ}{\mbox{$p_{\mathrm{T}}^{Z}$}}

\newcommand{\yZ}{\mbox{$y^{Z}$}}

\newcommand{\riso}{\mbox{$R_{\mathrm{iso}}$}}
\newcommand{\mTAA}{\mbox{$\langle T_{\mathrm{AA}}\rangle$}}

\usepackage{atlasphysics}

\begin{document}

\begin{frontmatter}



\dochead{}

\title{Measurement of the $W$, $Z$ and photon production in lead-lead collisions at \sqn = 2.76\,\TeV\ with the ATLAS detector}


\author{Alexander Milov, on behalf of the ATLAS Collaboration}

\address{Weizmann Institute of Science, 234 Herzl Street, Rehovot 76100, Israel}

\begin{abstract}
The ATLAS experiment measures yields of isolated photons and of $Z$ and $W$ bosons via leptonic decay modes in Pb+Pb collisions at \sqn=2.76 TeV. The data samples used in the analysis were obtained in the year 2010 and year 2011 LHC runs and correspond to 5 $\mu$b$^{-1}$ and 0.15 nb$^{-1}$ of integrated luminosity respectively.  The measured yields of all bosons are consistent with the scaling proportional to the number of nucleon-nucleon collisions. The transverse momentum distributions are measured for isolated photons and for $Z$ bosons for different centrality bins. The shapes of measured distributions do not change with centrality. The transverse momentum and rapidity distributions of $Z$ bosons are consistent in shape with {\sc Pythia} model.% and the second harmonic coefficient of the azimuthal distribution of $Z$ bosons with respect to the event plane is consistent with zero.
\end{abstract}


\end{frontmatter}

 \linenumbers




\section{Introduction}\label{sec:itro}
The study of the heavy ion (HI) collisions carried out by the experiments at the Relativistic Heavy Ion Collider (RHIC) at BNL, and the Large Hadron Collider (LHC) at CERN, demonstrated that the hot and dense matter produced in the interactions of heavy nuclei at high energy imposes a significant energy loss on the energetic color charge carriers penetrating such a medium~\cite{rhic_white}. An understanding of this phenomenon requires measuring the initial production rates of the particles in HI collisions before they lose energy. The best candidates to perform such measurements are particles which do not interact via the strong force, such as $W$, $Z$ bosons and photons. The PHENIX experiment at RHIC measured the properties of highly energetic photons~\cite{phenixphotons2}. The ATLAS and CMS experiments at LHC provided the first measurements of $Z$, $W$ and photons at the LHC energy \cite{Aad:2010aa, PhysRevLett.106.212301, ATLAS-CONF-2011-078, Robles:2011tq, Chatrchyan2012256}. Due to the complexity of these measurements and limited statistics of the first LHC \PbPb\ run, the results have relatively large uncertainties. Within their uncertainties the measurements show that in HI collisions the production rates of particles at high momentum transfer are proportional to average nuclei thickness function (\mTAA), commonly expressed also as mean number of binary collisions, \Ncoll, between the nucleons of the colliding nuclei. These two parameters are related as $\Ncoll = \mTAA\times\sigma_{pp}$, where $\sigma_{pp}$ is the total inelastic cross section of \pp\ interactions. The calculations of these parameters are carried out within the framework of the Glauber model~ \cite{Miller:2007ri,Alver:2008aq}. 

This proceeding presents the latest results from the ATLAS experiment at LHC on the measurement of $W$, $Z$ bosons and isolated photon production in \PbPb\ collisions at \sqn=2.76\,\TeV. The results measured in different centrality bins are compared to each other and also to predictions. 

\section{ATLAS detector}\label{sec:det}
The ATLAS detector~\cite{Aad:2008zzm} at the LHC covers nearly the entire solid angle around the collision point. It consists of an inner tracking detector surrounded by a thin superconducting solenoid, electromagnetic and hadronic calorimeters, and a muon spectrometer incorporating three large superconducting toroid magnet systems. 

The inner-detector system (ID) is immersed in a 2~T axial magnetic field and provides charged particle tracking in the pseudorapidity range $|\eta|<2.5$\footnote{The ATLAS reference system is a Cartesian right-handed coordinate system, with the nominal collision point at the origin. The anticlockwise beam direction defines the positive $z$-axis, while the positive $x$-axis is defined as pointing from the collision point to the center of the LHC ring and the positive $y$-axis points upwards. The azimuthal angle $\phi$ is measured around the beam axis, and the polar angle $\theta$ is measured with respect to the $z$-axis. Pseudorapidity is defined as $\eta = - \ln(\tan(\theta/2))$.}. The high-granularity silicon pixel (Pixel) detector covers the vertex region with three layers of pixels of $400\times50$\,$\mu$m$^2$ and it is followed by four layers of double sided silicon microstrip tracker (SCT) and the transition radiation tracker.

The calorimeter system covers the range $|\eta|< 4.9$. Within the region $|\eta|< 3.2$, electromagnetic calorimetry is provided by barrel and end-cap high-granularity lead-liquid argon (LAr) calorimeters, with an additional thin LAr presampler covering $|\eta| < 1.8$.

The muon spectrometer (MS) comprises separate trigger and high-precision tracking chambers measuring the deflection of muons in a magnetic field generated by superconducting air-core toroids. The precision chamber system covers the region $|\eta| < 2.7$ with three layers of monitored drift tubes~(MDT), complemented by cathode strip chambers~(CSC) in the innermost layer of the forward region, where the background is highest. The muon trigger system covers the range $|\eta| < 2.4$ with resistive plate chambers in the barrel, and thin gap chambers in the end-cap regions.

The two minimum-bias trigger scintillator (MBTS) counters, covering $2.1 < |\eta| < 3.9$ on each side of the nominal interaction point register charged particle hits with 16 scintillating pads on each side. Two zero-degree calorimeters (ZDC), each positioned at 140\,m from the collision point, detect neutrons and photons with $|\eta| > 8.3$.

\section{Analysis}\label{sec:ana}
The analysis of $W$ bosons uses 2010 LHC \PbPb\ collision data corresponding to an integrated luminosity of approximately 5\,\imb. The analysis of isolated photons and $Z$ bosons is based on the 2011 LHC \PbPb\ collision data and corresponds to an integrated luminosity of approximately 0.15\,\inb. Both data samples were obtained at the energy of \sqn=2.76\,\TeV.

\subsection{Event selection}\label{sec:evt}
All selected events are required to satisfy the Minimum Bias (MB) event selection. All events in 2010 run were taken with the MB trigger, requiring the coincidence of the ZDC signals on both sides, and the event vertex reconstructed offline. With increasing event rates in 2011 run, this condition was modified. On a trigger level the MB required transverse energy (\et) deposition of $\et>$50\,\GeV~ in the ATLAS calorimeter or a coincidence of the ZDC signals on both sides and a track in the ID system. In offline analysis all events were required to have an event vertex reconstructed by the ID tracking system and also an MBTS timing signal coincidence $|\Delta t|<3$\,ns. The total number of sampled events is different in different analyses. It varies from approximately $35\times10^{6}$ MB events selected for the analysis of $W$ in 2010 run to  $1.03\times10^{9}$ MB sampled events in the analysis of the \zee\ decay channel~\cite{ATLAS-CONF-2012-052} in 2011 data. %Based on previous studies~\cite{ATLAS:2011ag}, events satisfying the MB selection criteria, correspond to a fraction of the non-Coulomb inelastic \PbPb~ cross section equal to $\pm2$\%. 

\subsection{Event centrality}\label{sec:centr}
In HI collisions, ``centrality'' reflects the overlap volume of the two colliding nuclei, controlled by the impact parameter. That overlap volume is closely related to the average number of participants, \Npart, the nucleons which scatter inelastically in each nuclear collision and the number of binary collisions, \Ncoll, between the nucleons of the colliding nuclei.

The \PbPb~ collision centrality is measured using the sum of transverse energy (\sumet) deposited in the Forward Calorimeter (FCal) over the pseudorapidity range $3.1<|\eta|<4.9$ calibrated at the electromagnetic energy scale~\cite{ATLAS-CONF-2011-079}.  The fraction of events with more than one collision in the 2010 run was negligibly small, and in the 2011 run it was estimated not to exceed 0.05\% independent of centrality except for the very central collisions. A cut on the FCal energy of \sumet$<$3.8\,\TeV~ was applied to prevent pileup contamination in the analyzed data. 

\subsection{Reconstruction of the $W$ bosons}\label{sec:recoW}
\begin{figure}[htb]
\begin{center}
\includegraphics[width=0.38\textwidth]{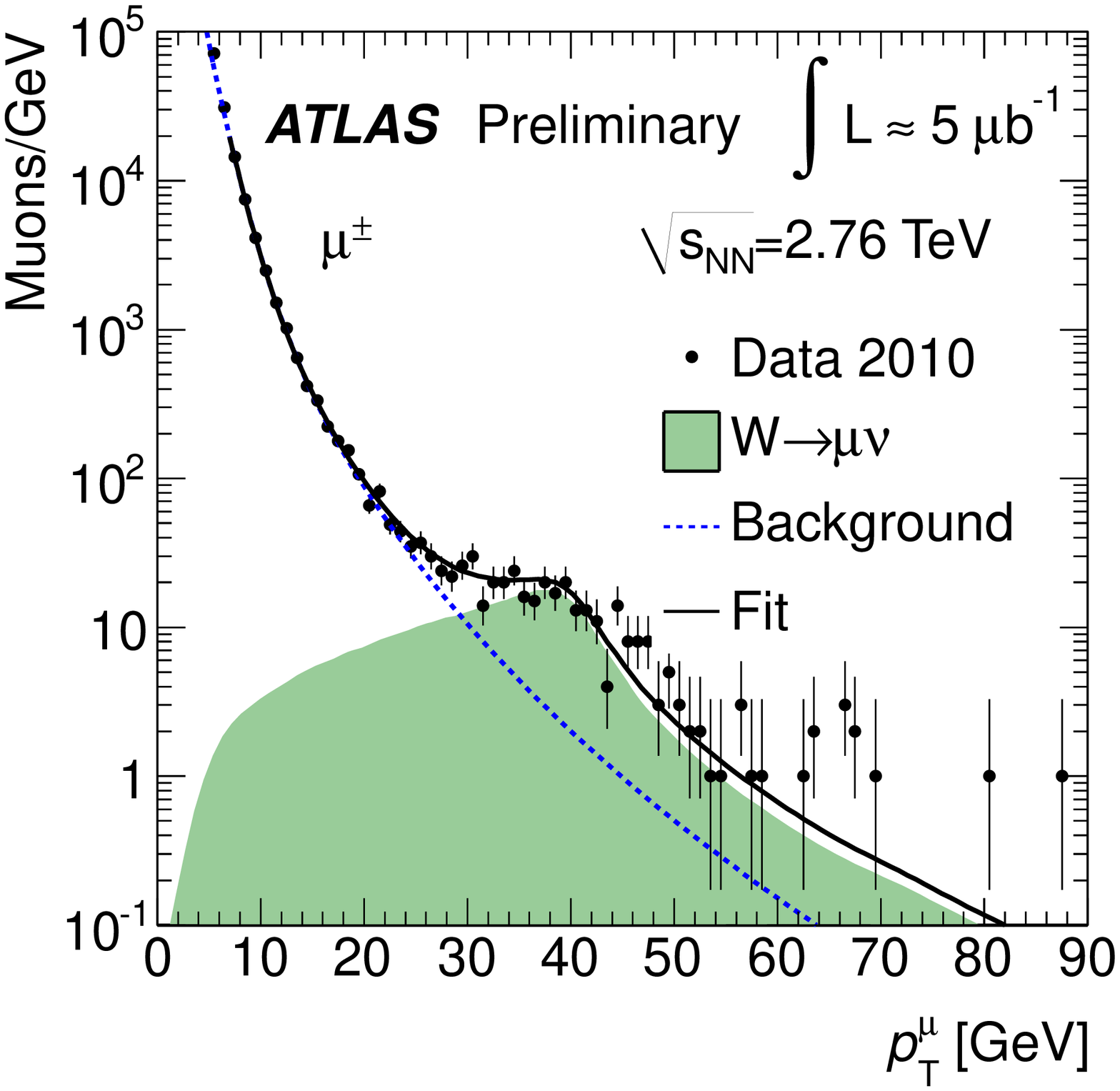}
\includegraphics[width=0.61\textwidth]{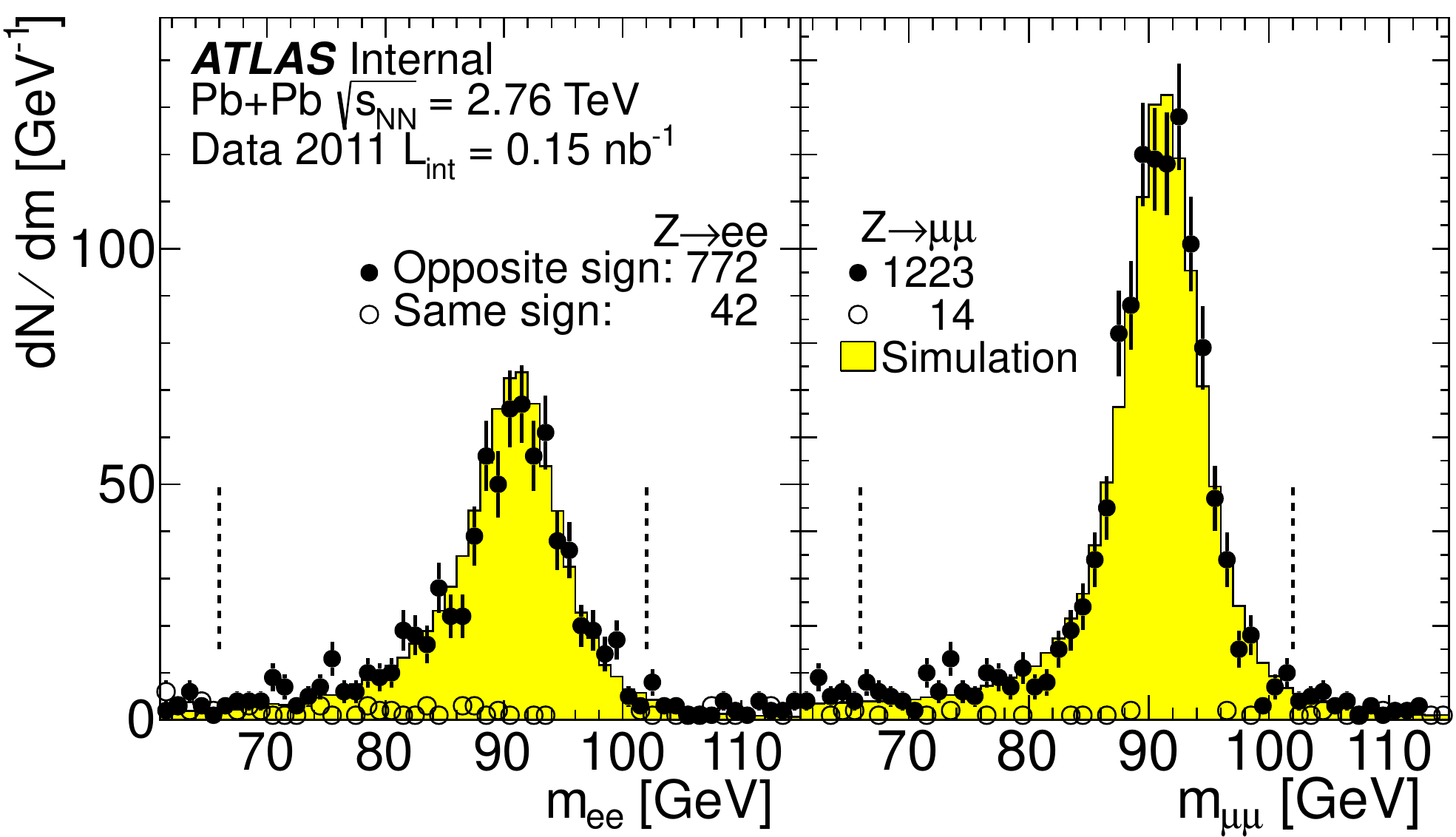}
\caption{Left: Uncorrected inclusive muon \pT\ spectrum from Ref.~\cite{ATLAS-CONF-2011-078}. The spectrum is fitted (solid line) with two components: signal $W\rightarrow\mu\nu$ (shaded area) simulated with {\sc Pythia} in \pp\ collisions, and a background parametrization (dashed line). The invariant mass distributions of \zee~ (middle) and \zmm~ (right) in data and MC integrated over momentum, rapidity, and centrality from Ref.~\cite{ATLAS-CONF-2012-052}. The MC is reweighted to match the centrality distribution in data and normalized in the region $66\,\GeV < m_{ll} < 102$\,\GeV~ ($l=e,\mu$) indicated by dashed lines. Numbers of counts given in the plot correspond to the same mass region.}
\label{fig:w_fit}
\end{center}
\end{figure}

Measurement of the $W$ bosons is based on reconstruction of inclusive muon spectrum. Measurements of the muon trajectories from both the ID and MS are combined, resulting in a relative momentum resolution ranging from about 2\% at low momentum up to about 3\% at transverse momentum (\pT) about 50\,\GeV. Muon tracks are required to have at least two hits in the Pixel detector, one of them in the first layer, and six or more hits in the SCT. To improve on the match of the muon spectrometer track and the inner detector the momentum measured in the muon spectrometer must be within 50\% of the corresponding measurement in the ID. Left panel of Fig.~\ref{fig:w_fit} shows the uncorrected muon spectrum obtained in all events in the 2010 run~\cite{ATLAS-CONF-2011-079}. 

The spectrum shows a steep power law fall up to \pT\ of 30\,\GeV\ where the presence of muons from the W decay appears prominently. The $W$ production yields are obtained performing fits to the muon transverse momentum spectra using two input shapes. The first shape, shown as shaded area in the left panel of Fig.~\ref{fig:w_fit} describes the muon \pT\ distribution from $W^{\pm}\rightarrow\mu^{\pm}\nu$, and a second describes the background of exponential shape, obtained from studies of $c\bar{c}\rightarrow\mu+X$ and $b\bar{b}\rightarrow\mu+X$ decays in \pp\ simulations and described in~\cite{ATLAS-CONF-2011-078}.

For the present analysis, only muons with $\pT>7$\,\GeV\ are considered, and also muon pairs forming the invariant mass $m_{\mu\mu}>66$\,\GeV\ are removed from the analysis of the $W$ decays to reduce the background coming from $Z$. After applying cuts, the total number of $W^{\pm}$ candidates extracted from the event sample by fitting the templates is $399^{+36}_{-38}$.

\subsection{Reconstruction of the isolated photons}\label{sec:recoG}

Photon candidates were identified at the first trigger level (L1) by a cluster formed by $(\Delta\phi\times\Delta\eta)=0.1\times0.1$ trigger towers of the electromagnetic part of the calorimeter, covering a pseudorapidity range $|\eta|<2.5$. A cluster transverse energy estimated at L1 was required to exceed \et=16\,\GeV.  The trigger becomes fully efficient to the photons with energy exceeding 20\,\GeV. 

In order to reconstruct photons and electrons in the context of a heavy ion collision, the large background from the underlying event (UE) is subtracted from each event. This is performed in the same way as in the heavy ion jet reconstruction, explained in detail in Ref.~\cite{jet2}. The UE subtraction takes into account the average energy density in the calorimeter modulated by the flow effects in the regions excluding jet candidates.

As described in Ref.~\cite{ATLAS-CONF-2012-051}, after applying the shower shape cuts, but before applying isolation cuts  there are 6435 photon candidates with $\pT > 45$\,\GeV\ and within $|\eta|<1.3$ the (0-80)\% centrality sample. Additional kinematic constraints on photon candidates are used to restrict the analysis to the region with less material in front of the calorimeter and to improve the performance of the photon identification. 

\begin{figure}[htb]
\begin{center}
\includegraphics[width=0.38\textwidth]{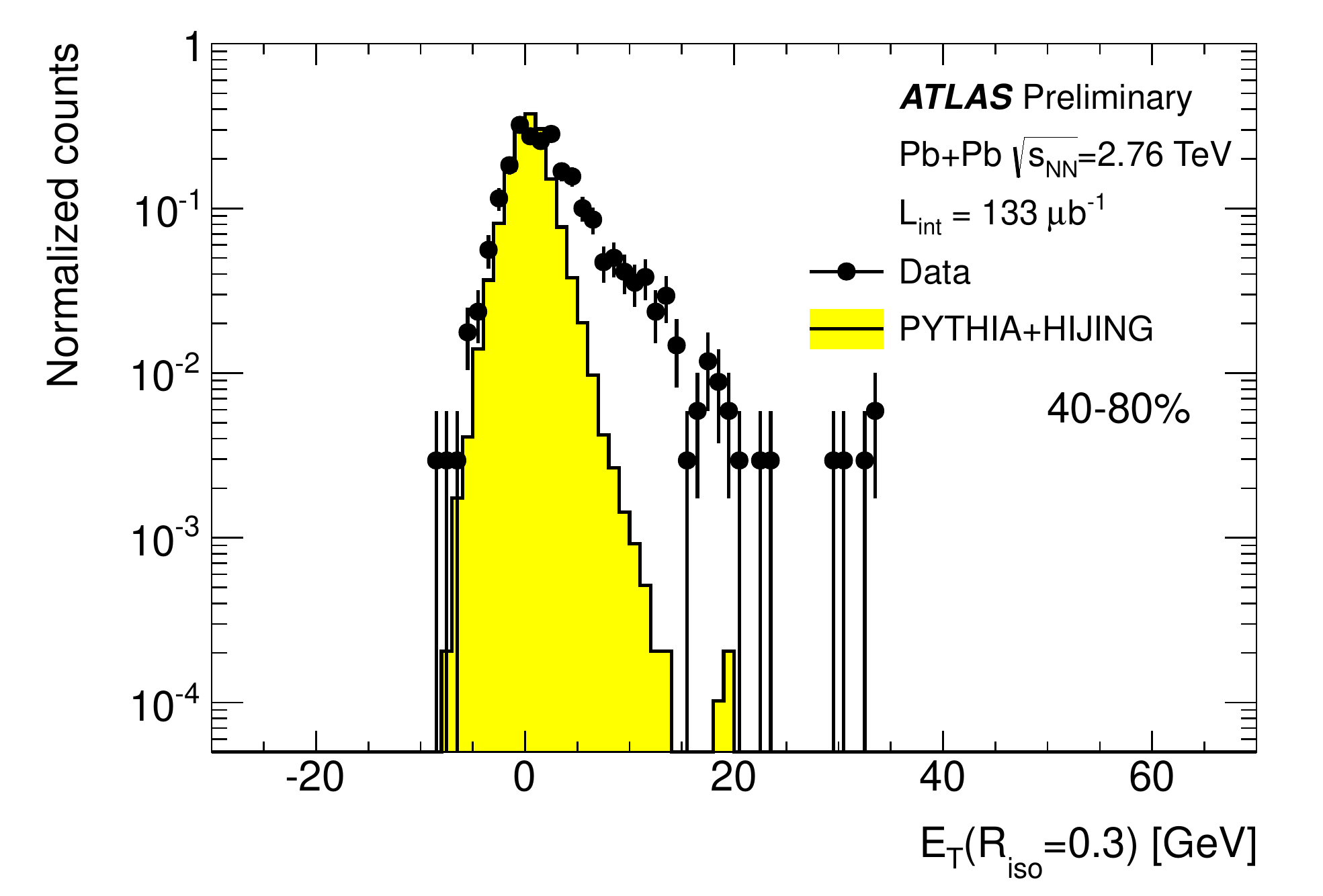}
\includegraphics[width=0.38\textwidth]{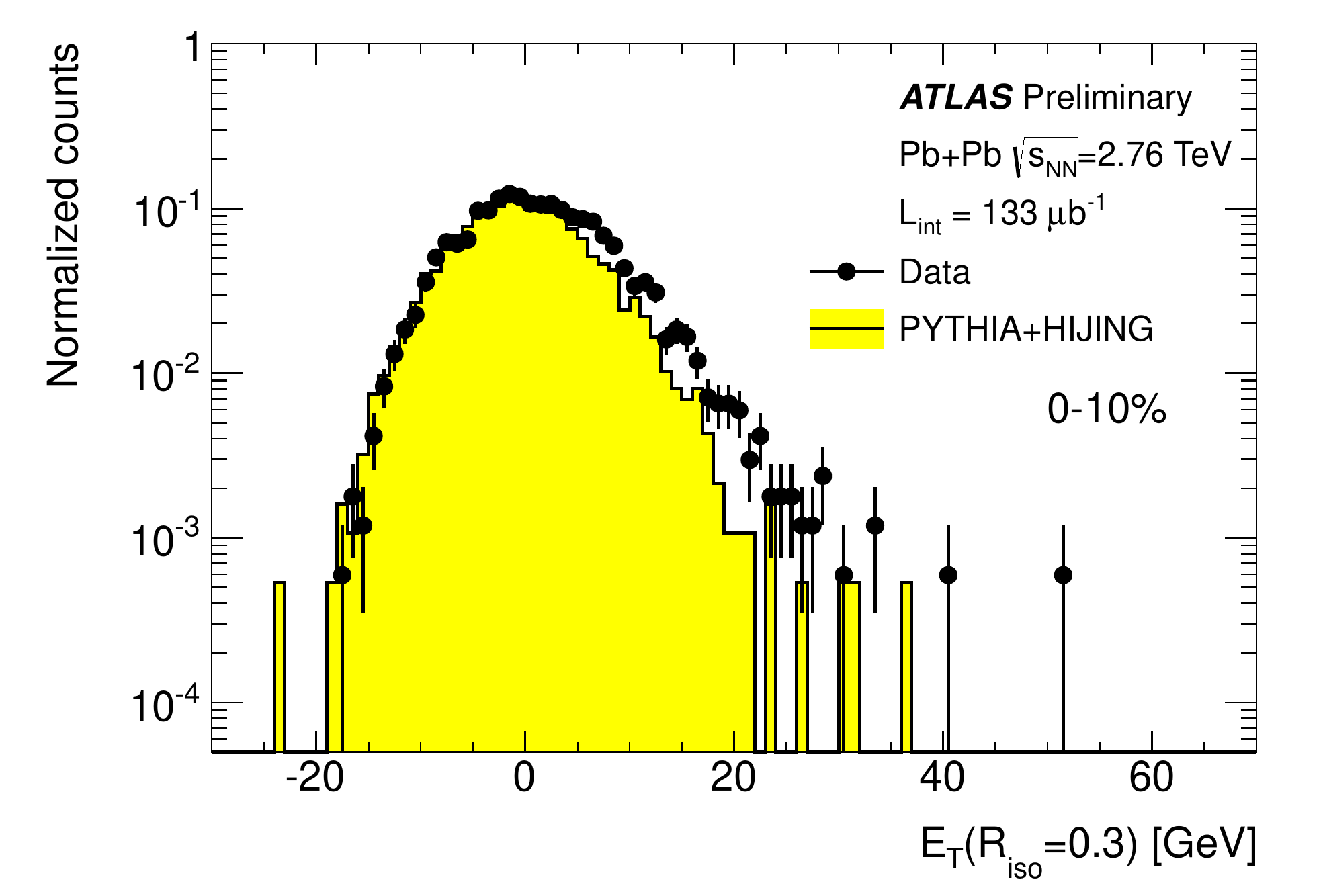}
\caption{Distributions of photon isolation energy in a $\riso = 0.3$ cone for the three centrality bins in data (black points) and for MC (yellow histogram), normalized for negative \et\ values from Ref.~\cite{ATLAS-CONF-2012-051}. The differences at large values of \et\ can be attributed to the presence of jet contamination in the data, which is not present in the MC sample.}
\label{fig:e_iso}
\end{center}
\end{figure}

The isolation energy is the sum of transverse energies in calorimeter cells within a cone or a radius $\riso=\sqrt{\Delta\eta^{2}+\Delta\phi^{2}}$ around the photon direction, excluding the cells associated to the photon itself. Figure~\ref{fig:e_iso} shows the distributions of \et\ for $\riso = 0.3$ as a function of collision centrality. 
The data and MC distributions shown in the figure grow noticeably wider with increasing centrality, as the UE subtraction is only able to subtract the mean energy in an interval. The effect is reproduced by the MC, except for a contribution from di-jets which are not present in simulation. An isolation cut of 6\,\GeV\ was used in the analysis. %, removing reconstructed jets, so local fluctuations are still preserved. Furthermore, in the data, an enhancement on the ET(R_{iso}=0.3)> 0 side of these distributions is expected from di-jet background. Still it is observed that the isolation distributions in data and simulation vary in a similar fashion, especially in the region of negative isolation energy.
The absolute energy scale of the photons derived from the {\sc Pythia} events embedded into {\sc Hijing} UE is withing 1.5\% of the generated value. The energy resolution is better than 3\% in the most central collisions. 

The photon yields are extracted using the double sideband technique which divides photon candidates according to their quality and isolation energy as described in~\cite{ATLAS-CONF-2012-051}. This technique provides an accurate estimation of residual background in the isolated photon sample and signal leakage in the background region.

\subsection{Reconstruction of the $Z$ bosons}\label{sec:recoZ}
The $Z$ bosons are reconstructed via $\mu\mu$ and $ee$ decay channels. A data sample for this study was collected by the ATLAS three-level trigger system~\cite{Aad:2012xs} which selected events with high-\pt\ muon and electron candidates in 2011 data sample.

%\subsubsection{\zmm}
High-\pt~ muon candidates were selected using all three levels of the ATLAS trigger system. The L1 muon trigger searches for patterns of hits consistent with muons at certain \pt~ in the trigger chambers within $|\eta|<2.4$. If a muon \pt~ estimate exceeds 4\,\GeV, the event is accepted for further processing at the High Level Trigger~(HLT). The L1 muon algorithm also identifies Regions of Interest~(RoI) within the detector to be investigated by the HLT.  At the HLT, each muon candidate's track parameters are recalculated by including the precision data from the MDT or CSC in the RoI defined by the previous trigger level. Muon candidates are reconstructed either solely from the MS or using combined data from the MS and ID systems. In addition to the events selected using the RoI-based muon trigger, an event filtering trigger was used to identify muons with $\pt>10$\,\GeV. The filtering was done for all events satisfying a L1 requirement of the ZDC firing on both sides or an $\et>10$\,\GeV~ deposition in the calorimeter.

In the \zmm~ analysis, single muons may be reconstructed with varying levels of quality\cite{ATLAS-CONF-2010-036}. High quality muons are reconstructed in the MS and ID subsystems and have a consistent angular measurement in both subdetectors as well as a good match to the primary vertex. At least one muon in a pair, matched to the trigger, is required to be of such quality. If the second muon in the pair is fully reconstructed as a track by both subsystems, the minimum \pt~ threshold is set to 10\,\GeV~ on both muons. If the second muon fails this condition, both muons are required to satisfy \pt$>$20\,\GeV. An invariant mass window of $66\,\GeV < m_{\mu\mu}<102$\,\GeV~ is used to define oppositely charged muon pairs as $Z$ boson candidates and same-sign charged pairs as a background estimate. In total, 1223 opposite-sign candidates and 14 same-sign candidates reconstructed in the \zmm~ channel are shown in the right panel of Fig.~\ref{fig:w_fit}.

%\subsubsection{\zee}

Electron candidates are reconstructed at L1 requiring \et\ to exceed 14\,\GeV.  The UE subtraction is performed for photon clusters in the same way as for the photon candidates. 

%Electron candidates were identified at the first trigger level (L1) by a cluster formed by $(\Delta\phi\times\Delta\eta)=0.1\times0.1$ trigger towers of the electromagnetic part of the calorimeter, covering a pseudorapidity range $|\eta|<2.5$. A cluster transverse energy estimated at L1 was required to exceed \et=14\,\GeV.

%In order to reconstruct photons and electrons in the context of a heavy ion collision, the large background from the underlying event (UE) is subtracted from each event. This is performed in the same way as in the heavy ion jet reconstruction, explained in detail in Ref.~\cite{jet2}. The UE subtraction takes into account the average energy density in the calorimeter modulated by the flow effects in the regions excluding jet candidates.

For the \zee~ analysis, electron candidates are formed using the standard ATLAS reconstruction algorithm, requiring the matching of an ID track to a LAr cluster.  Electron selection is limited to $|\eta|<2.5$ and both electrons are required to have \et$>$20\,\GeV.  The standard electron identification cuts~\cite{Aad:2011mk} used in the \pp~ environment are not suited to the \PbPb~ environment  due to the amount of energy from the underlying event deposited in the calorimeter. To address this, a different set of cuts was developed to accommodate the modification of the calorimeter variables by the presence of the underlying event. The electron identification cuts used were based on the energy balance between the track and EM cluster ($E/p$) and shower shape variables.  Since almost all electrons with \et$>$20\,\GeV~ are reconstructed as jets by the heavy-ion anti-$k_{\mathrm{t}}$ R=0.2 jet finding algorithm~\cite{jet2}, a further cut was made requiring that the \et~ of the electron be at least 53\% of the jet \et.  %In addition to the cuts, the underlying event was estimated (following~\cite{jet2}) and subtracted  on an electron by electron basis to recover the proper electron energy.

%Electron reconstruction and identification efficiency is evaluated using electrons from $7\times10^{5}$ {\sc Pythia}~\cite{Sjostrand:2006za} \zee~ events  with $66\,\GeV < m_{Z} < 116$\,\GeV~ and $|\yZ|<2.5$ embedded into \PbPb~ events generated by the {\sc Hijing}~\cite{PhysRevD-44-3501} event generator. The response of the ATLAS detector to the generated particles is modeled using GEANT4~\cite{Agostinelli2003250}. The identification and reconstruction efficiency for electrons of \et$>$20\,\GeV, after all cuts, ranges from 72\% in the most central collisions to 76\% in peripheral collisions.

For the \zee~ analysis all electrons found in triggered events were paired with each other, requiring that at least one electron in the pair was matched to a trigger object. The opposite-sign pairs within an invariant mass satisfying $66\,\GeV < m_{ee} < 102$\,\GeV~ are accepted as signal Z candidates and the same charged-sign pairs in this window are taken as an estimate of the combinatorial background. In total, 772 opposite-sign pairs and 42 same-sign pairs reconstructed in the \zee~ decay channel are shown in Fig.~\ref{fig:w_fit}.

\subsection{Corrections and uncertainties}\label{sec:corr}
The $W$ boson is measured using the MB event sample, for the measurement of $Z$ boson and isolated photons, measured with triggered sample, correction for finite efficiency of the trigger system is applied as explained in sec.~\ref{sec:recoZ}. In \zee\ channel and in the measurement of isolated photons these corrections are significantly smaller than 1\%.

Reconstruction and identification efficiency of $Z$ bosons was evaluated using $7\times10^{5}$ \zee\ events and $5.3\times10^{5}$ \Zmm\ events with $66\,\GeV < m_{Z} < 116$\,\GeV~ and $|\yZ|<2.5$ produced by {\sc Pythia} event generator~\cite{Sjostrand:2006za} and embedded into \PbPb~ events generated by the {\sc Hijing}~\cite{PhysRevD-44-3501} event generator. The response of the ATLAS detector to the generated particles is modeled using GEANT4~\cite{Agostinelli2003250}. 

The $Z$ boson distributions are corrected for the losses due to detector acceptance and analysis cuts in the bins of \pT, $\eta$ and centrality. The correction is larger for the \zee\ channel, as the electron losses are higher. At the very edge of the acceptance region the \zee\ reconstruction efficiency drops to 10\% from more than 50\%  at midrapidity. The corresponding numbers for the \zmm\ measurements are 45\% and 70\% respectively. 

The main source of systematic uncertainties in the measurement of isolated photons are the photon quality cuts and photon isolation criteria. The uncertainties reach 20\% for each. In the measurement of the $Z$ bosons the main contributions in both decay channels are from the uncertainties of efficiency corrections reaching up to 8\%. Because both channels have different sources of systematic uncertainties they can be averaged taking into account the statistical and systematic uncertainties. Finally, the uncertainties on the \Ncoll\ reach more than 10\% and they play dominant role in several results.

\subsection{Results}\label{sec:res}
\begin{figure}[htb]
\begin{center}
\includegraphics[width=0.69\textwidth]{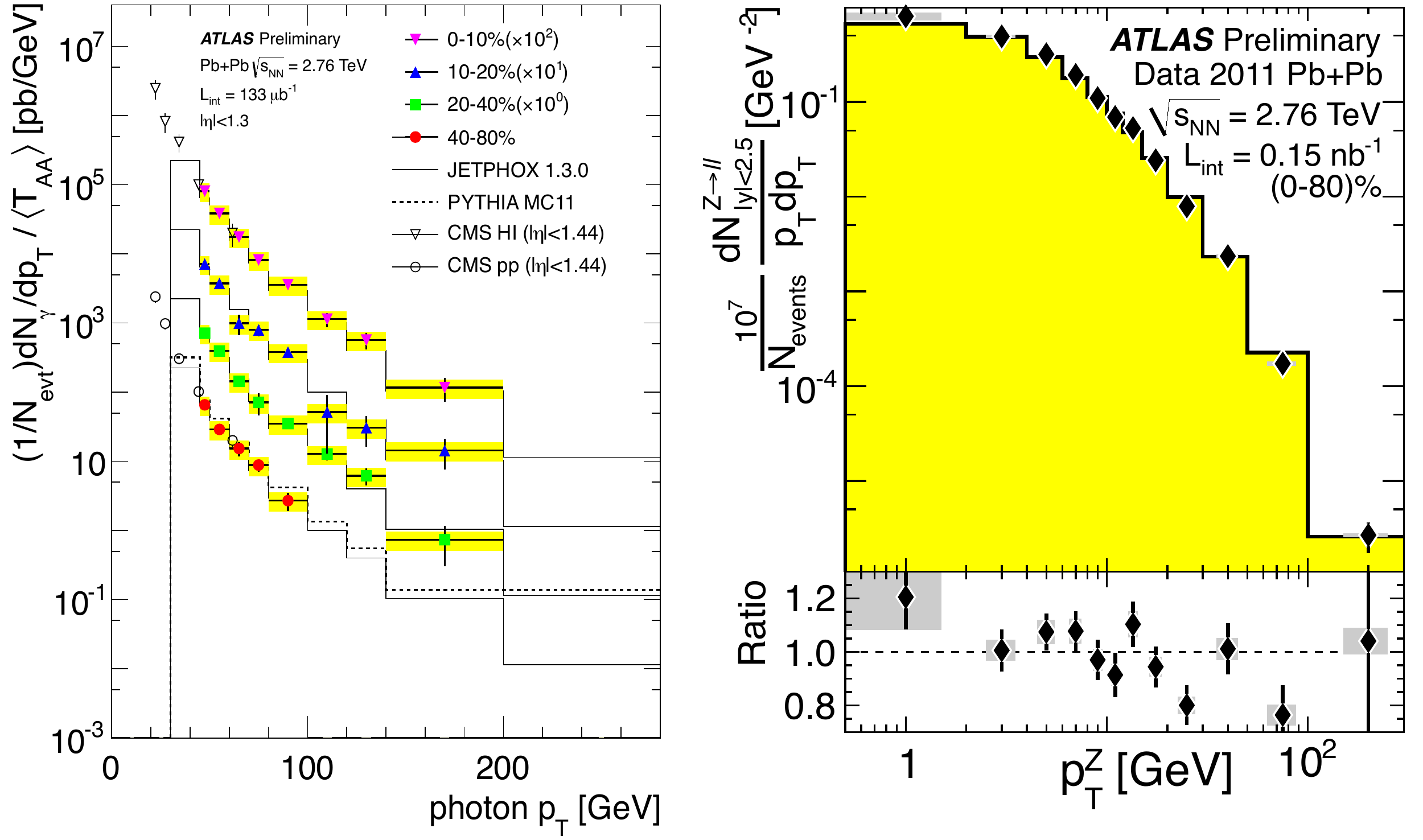}
\caption{Left: Efficiency corrected yields of prompt photons in $|\eta|<1.3$ in the isolation cone radius $\riso = 0.3$ and isolation energy of 6\,\GeV, scaled by \mTAA\ from Ref.~\cite{ATLAS-CONF-2012-051}. Statistical errors are shown by the error bars. Systematic uncertainties on the photon yields are combined and shown by the yellow bands. The scale uncertainties due only to \mTAA. The solid lines are {\sc Jetphox} calculation of photon yields and dashed line is the {\sc Pythia}. Corrected \ptZ\ (middle) and rapidity (right) distributions of the measured $Z\rightarrow l^{+}l^{-}$ ($l=e,\mu$) data compared to {\sc Pythia} predictions normalised by area from Ref.~\cite{ATLAS-CONF-2012-052}. The error bars represent statistical uncertainties, and the filled bands represent systematic uncertainties. The lower panes display the ratio of data over {\sc Pythia}.}
\label{fig:pt_rap}
\end{center}
\end{figure}

The fully corrected \pT\ distribution of isolated photons measured by the ATLAS experiment is shown in the left panel of Fig.~\ref{fig:pt_rap}.
The CMS Collaboration measured production of isolated photons~\cite{Chatrchyan2012256} in heavy ion collisions at \sqn=2.76\,\TeV\ in the rapidity region $|\eta|<1.44$ and an isolation condition of at most 5\,\GeV\ transverse energy in a cone of radius $\riso = 0.4$. The measurement extends to 80\,\GeV. The CMS  (0-10\%) data are superimposed on the most central bin as well as the CMS \pp\ data are superimposed on the most peripheral bin. Within the range of both measurements the results are consistent.

For both \zee~ and \zmm~ analyses, correction factors to account for the efficiency and detector resolution within the selected acceptance based on the simulation are calculated differentially in event centrality, \ptZ, and \yZ.  In each decay channel, the correction factor is applied and the background, estimated by the same-sign charged pairs, is subtracted.  The two decay channel measurements are then combined with weights set by their respective uncertainties. The fully corrected \ptZ\ distribution for the $Z$ boson is shown in Fig.~\ref{fig:pt_rap}. The data distributions plotted in (0-80)\% centrality agree in shape with the {\sc Pythia} simulations of $Z$ boson production in \pp~ collisions.

%%%%%%%%%%%%%%%%%%%%%%%%%%%%%%%%%%%%%
The nuclear modification factor, \rpc, normalized to the statistically-robust central collisions bin can be defined as:
\begin{equation}
\rpc = \frac{\Ncoll (C)}{\Ncoll (P)} \frac{(1/N_{\mathrm{evt,P}}) \mathrm{d}^2 N_{\mathrm{P}} / \mathrm{d}y \mathrm{d}\pT}{ (1/N_{\mathrm{evt,C}}) \mathrm{d}^2 N_{\mathrm{C}} / \mathrm{d}y \mathrm{d}\pT}
\label{eq:rcp}
\end{equation}
where \Ncoll (P) and \Ncoll (C) are the number of binary nucleon-nucleon collisions calculated for peripheral and central collisions respectively, and $(1/N_{\mathrm{evt,C}}) \mathrm{d}^2 N_{\mathrm{C}}/\mathrm{d}y d\pT $ and $(1/N_{\mathrm{evt,P}}) \mathrm{d}^2 N_{\mathrm{P}}/\mathrm{d}y d\pT $ are the differential yields of $Z$ per-event for central and peripheral collisions respectively.

\begin{figure}[b!]
\begin{center}
\includegraphics[width=0.70\textwidth]{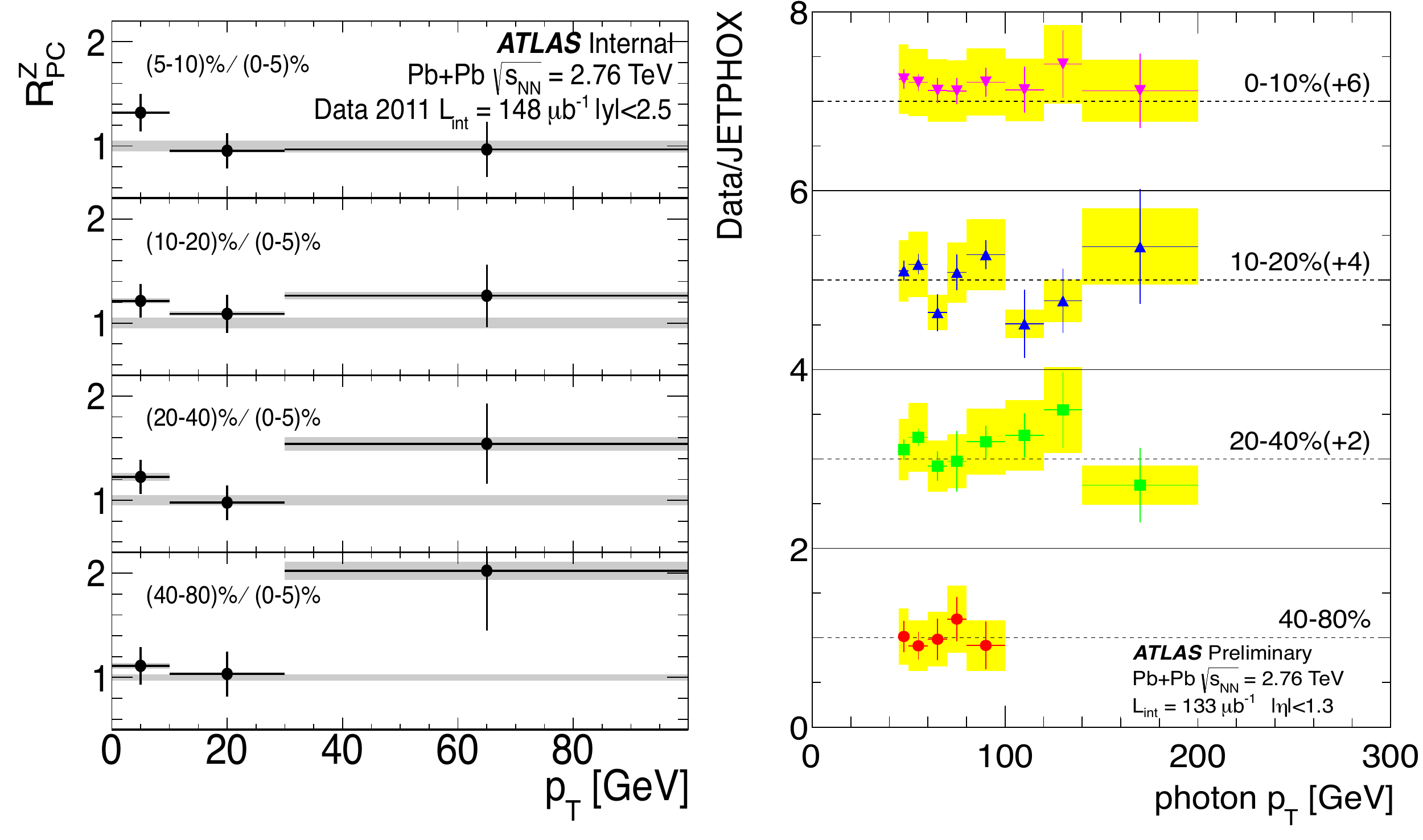}
\caption{Left: the \rpc\ of $Z$ boson for four centrality bins. The error bars show the statistical uncertainties, and the shaded areas at the point location, the residual systematic uncertainty from Ref.~\cite{ATLAS-CONF-2012-052}. Additional uncertainty, coming from the \Ncoll (P) to \Ncoll (C) ratio is the same for all points in the panel and is shown centered about unity. Right: photon yields in $|\eta|<1.3$ in isolation cone radius $\riso = 0.3$ and isolation energy of 6 GeV, divided by {\sc Jetphox} 1.3 predictions from Ref.~\cite{ATLAS-CONF-2012-051}. Statistical errors are shown by the error bars. Systematic uncertainties on the photon yields are combined and shown by the yellow bands.}
\label{fig:rpc}
\end{center}
\end{figure}

The nuclear modification factor for the $Z$ boson as a function of \ptZ\ is shown in the left panel of Fig.~\ref{fig:rpc} 
for several centrality bins. The systematic uncertainties largely cancel out in the ratio, however the statistical uncertainties remain large, especially for the most peripheral (40-80)\% bin. The statistical uncertainties are correlated in different panels of the figure, because all ratios use the same (0-5)\% centrality data as the denominator.  The right panel of Fig.~\ref{fig:rpc} shows the analogue results for isolated photons presented as ratios of the photon spectra divided by the results of the predictions from the {\sc Jetphox} and by the \mTAA.  Within the uncertainties, the \rpc\ of $Z$ bosons, and the ratio of photon production to the prediction values are consistent with unity. 

%\begin{wrapfigure}{l}{7cm}[htb]
\begin{figure}[htb]
\begin{center}
\includegraphics[width=0.35\textwidth]{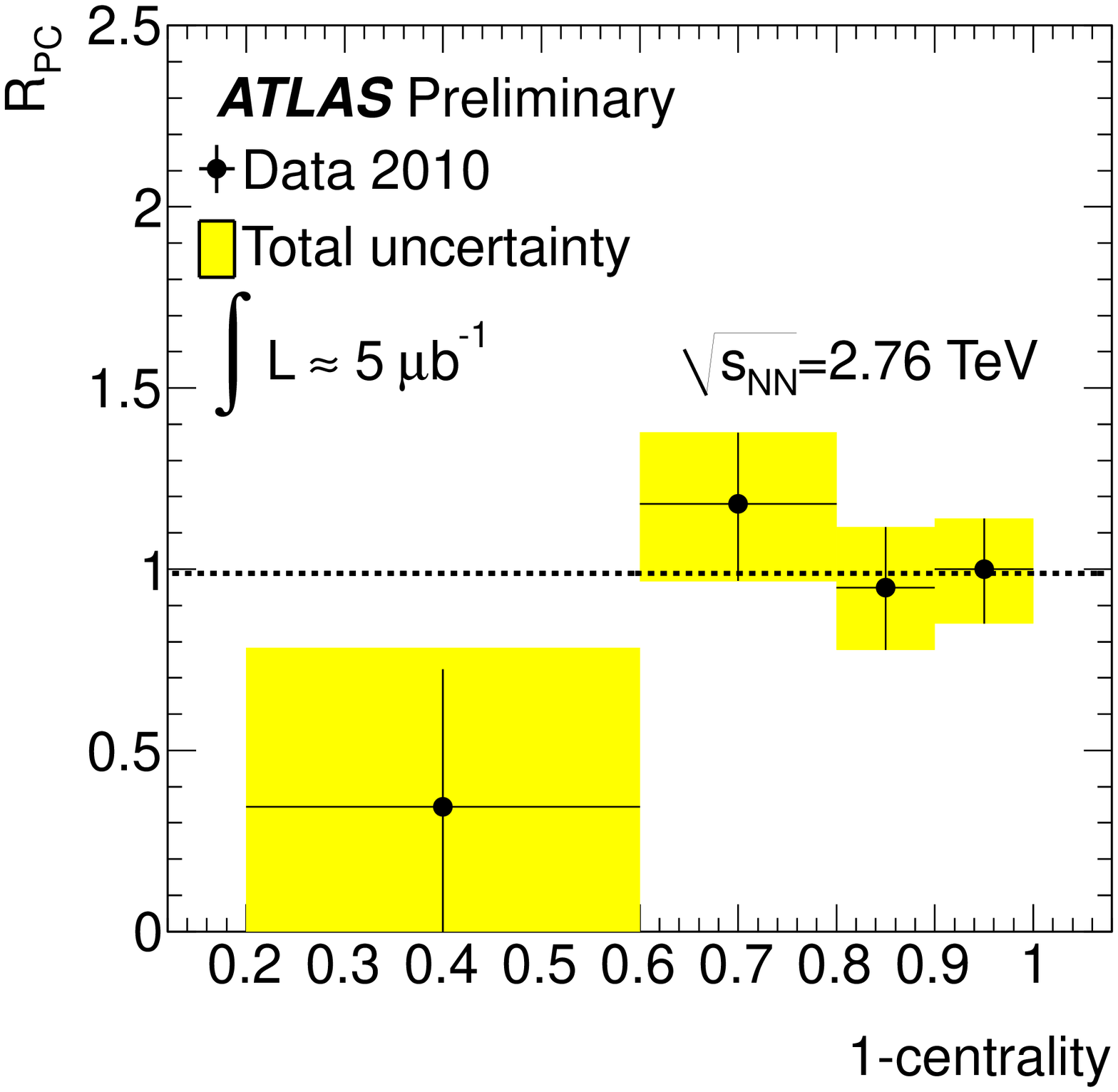}
\caption{$W$ boson \rpc\ as a function of collision centrality from Ref.~\cite{ATLAS-CONF-2011-078}. The error bars are uncertainties from the fit method and statistical uncertainties. Yellow bands include uncertainties from the number of binary collisions. The dashed line is the result of a flat line fit.}
\label{fig:w-scaling}
\end{center}
\end{figure}
%\end{wrapfigure}
Figure~\ref{fig:w-scaling} shows the \rpc\ of the $W$ boson as a function of centrality. Within the uncertainties, which are mainly statistical it remains consistent with unity.

\begin{figure}[h!]
\begin{center}
\includegraphics[width=0.70\textwidth]{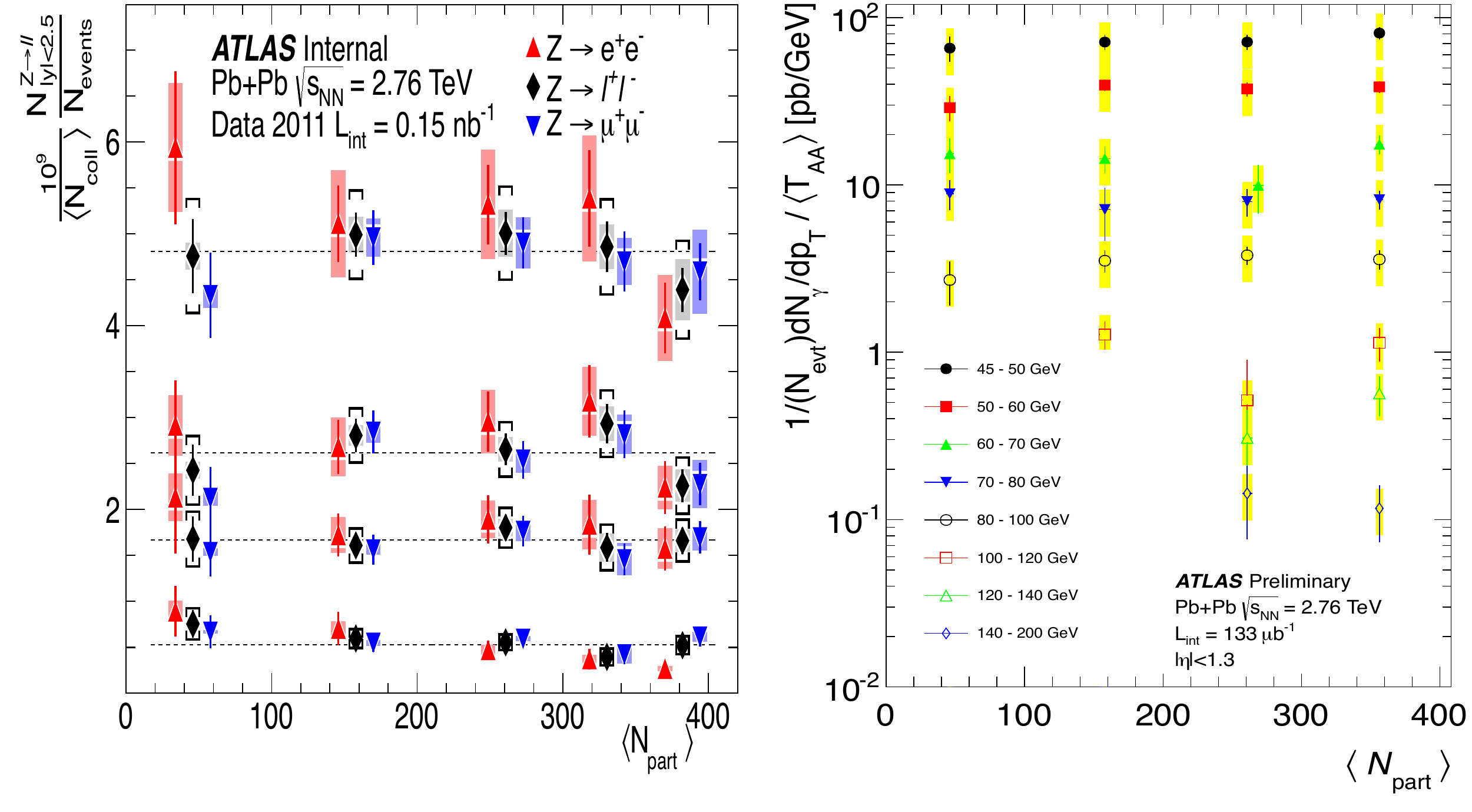}
\caption{Left: centrality dependence of $Z$ boson yields divided by \Ncoll, measured in $|\yZ|<2.5$. Results for $ee$ (upward pointing triangles) and $\mu\mu$ (downward pointing triangles) channels are shifted left and right respectively (for visibility) from their weighted average (diamonds) which is plotted at the nominal \Npart~ value from Ref.~\cite{ATLAS-CONF-2012-052}. The statistical (bars) and systematic (shaded bands) uncertainties are calculated using the appropriately weighted average of the two contributing sources. Brackets show the combined uncertainty including the uncertainty on \Ncoll. The dashed lines are constant fits to the combined results. Right: centrality dependence of the photon yield per event in each \pT\ bin, scaled by the average nuclear thickness function \mTAA\ for that centrality interval from Ref.~\cite{ATLAS-CONF-2012-051}. The horizontal axis is the average number of participants \Npart\ for each selected centrality interval. Statistical errors are shown by the error bars. Systematic uncertainties on the photon yields are combined and shown by the yellow bands.}
\label{fig:scaling}
\end{center}
\end{figure}

The left panel of Fig.~\ref{fig:scaling} shows the yields of isolated photon scaled by \mTAA\ as a function of \Npart. 
The $Z$ boson yields per-event, divided by \Ncoll, are shown in the right panel of Fig.~\ref{fig:scaling} for \zee\ and \zmm\ independently. The figure demonstrates that the two channels are consistent within their uncertainties and can be combined together. The combined results are also shown in the right panel. All values are integrated over the pseudorapidity range in which the corresponding particles are measured. The results for photons and $Z$ bosons are presented in several \pT\ slices and for $W$ and $Z$ bosons integrated over all \pT. Figure~\ref{fig:scaling} shows that the  yields of all measured bosons scale with the number of binary collisions \Ncoll\ (or nuclear thickness function \mTAA) calculated with the Glauber model.

%%%%%%%%%%%%%%%%%%%%%%%%%%%%%%%%%%%%%

%The amplitude of the second Fourier harmonic of the $Z$ boson azimuthal emission angle with respect to the event plane  %elliptic flow coefficient, ($v_2$) is shown in Fig.~\ref{fig:flow} for the centrality region (0-80)\%.
%\begin{figure}[h!]
%\begin{center}
%\includegraphics[width=0.6\textwidth]{fig_pool/fig5_final.pdf}
%\caption{The second harmonic $v_2$ as a function of $|\yZ|$ (left), \ptZ~ (center) and \Npart~ (right) from Ref.~\cite{ATLAS-CONF-2012-052}. The shaded areas show the systematic uncertainties of the measurements. The dashed lines show constant fits to the $v_2$ values, weighting the measurements by statistical uncertainties only.}
%\label{fig:flow}
%\end{center}
%\end{figure}
%The $v_2$ of the $Z$ boson anisotropic flow is measured to be $v_2=-0.015 \pm 0.018$(stat.)$\pm 0.014$(sys.), and is consistent with zero in rapidity, \ptZ, and \Npart.  This observation is an independent measurement consistent with binary collision scaling.

\section{Conclusions}\label{sec:con}
The ATLAS experiment measured the $W$, $Z$ boson and isolated photon production in \PbPb~ collisions at \sqn = 2.76\,\TeV~ using data collected in the 2010 and 2011 LHC physics runs. Isolated photons are reconstructed in the range $\pT = 45-200$\,\GeV\ and for $|\eta| < 1.3$. The $W$ boson yields are measured via single muons registered in the range $|\eta| < 2.5$ and with $\pT>7$\,\GeV. $Z$ bosons yields are fully corrected to $|\yZ|<2.5$ for the mass region $66\,\GeV < m_{Z} < 116$\,\GeV. Analyzed \zee\ and \zmm\ decay modes produce consistent results. The measured distributions of isolated photons agree in shape and in yield with the {\sc Jetphox} predictions calculated without taking into account nuclear modification. The shape of the \pT\ and $y$ distributions for $Z$ are consistent with the prediction of {\sc Pythia} generator. The total yields of all measured bosons scale proportional to the nuclear thickness function, \mTAA\ (or \Ncoll). The Nuclear Modification Factor \rpc\ for $Z$ and generator normalized for photons agree with unity as a function of \pT. %The second coefficient of azimuthal anisotropy of the $Z$ boson production is consistent with zero.

\section{Acknowledgments} 
This research is supported by FP7-PEOPLE-IRG (grant 710398), Minerva Foundation (grant 7105690) and by the Israel Science Foundation (grant 710743).

\end{document}